# Nanoscale origins of the damage tolerance of the high-entropy alloy CrMnFeCoNi

ZiJiao Zhang[1], M.M. Mao[1], Jiangwei Wang[2], Bernd Gludovatz[3], Ze Zhang[1], Scott X. Mao[1,2], Easo P. George[4], Qian Yu[1] & Robert O. Ritchie[3,5]

Damage tolerance can be an elusive characteristic of structural materials requiring both high strength and ductility, properties that are often mutually exclusive. High-entropy alloys are of interest in this regard. Specifically, the single-phase CrMnFeCoNi alloy displays tensile strength levels of ~1 GPa, excellent ductility (~60–70%) and exceptional fracture toughness ($K_{JIc} > 200$ MPa$\sqrt{m}$). Here through the use of *in situ* straining in an aberration-corrected transmission electron microscope, we report on the salient atomistic to micro-scale mechanisms underlying the origin of these properties. We identify a synergy of multiple deformation mechanisms, rarely achieved in metallic alloys, which generates high strength, work hardening and ductility, including the easy motion of Shockley partials, their interactions to form stacking-fault parallelepipeds, and arrest at planar slip bands of undissociated dislocations. We further show that crack propagation is impeded by twinned, nanoscale bridges that form between the near-tip crack faces and delay fracture by shielding the crack tip.

[1] Department of Materials Science & Engineering, Center of Electron Microscopy and State Key Laboratory of Silicon Materials, Zhejiang University, Hangzhou 310027, China. [2] Department of Mechanical Engineering & Materials Science, University of Pittsburgh, Pittsburgh, Pennsylvania 15261, USA. [3] Materials Sciences Division, Lawrence Berkeley National Laboratory, Berkeley, California 94720, USA. [4] Institute for Materials, Ruhr University, 44801 Bochum, Germany. [5] Department of Materials Science & Engineering, University of California, Berkeley, California 94720, USA. Correspondence and requests for materials should be addressed to Q.Y. (email: qyuzju@gmail.com) or to R.O.R. (email: roritchie@lbl.gov).

Damage tolerance is arguably the most important mechanical property of structural materials for many design applications, as it defines the combination of strength and toughness, that is, the ability of a material to resist fracture in the inevitable presence of flaws[1–5]. However, in many materials, particularly metallic alloys, toughness comes at the expense of strength[5]. Intrinsic toughness is promoted by extensive plastic deformation, which requires the easy motion and multiplication of dislocations (that can compromise strength), yet contradictorily requires significant hardening to generate strength and ductility by resisting dislocation motion and delaying plastic (necking) instabilities[2,5,6]. Despite this 'conflict', the ability to resist cracking in metals and alloys under mechanical loading can be affected by the prevailing modes of plastic deformation and microstructure at the crack tip. Generally, dislocation glide is the primary deformation mechanism at low homologous temperatures in crystalline materials where dislocations move by locally breaking and reforming crystal bonds. In traditional metallurgy, various alloying elements are usually added to the principal element to develop required mechanical properties[7,8]. The solute atoms locally introduce different crystal bonds with varied strengths and also cause local lattice distortion, thereby offering resistance to dislocation motion; as a result, strength is generally enhanced but ductility can be reduced.

Interestingly, a prominent member of the new class of high-entropy alloys (HEAs), CrMnFeCoNi, which forms a face-centered cubic (fcc) solid solution[1,9–12], has been found to display excellent damage tolerance with tensile strengths of ~1 GPa (refs 1,10,11), ductilities of ~60–70% (refs 1,10,11), and fracture toughness values exceeding 200 MPa$\sqrt{m}$ (ref. 1), properties that are comparable to the very best cryogenic steels, specifically austenitic stainless steels[13] and high-Ni steels[14–17]. At relatively low strains (~2%), deformation in CrMnFeCoNi occurs on the normal fcc {111}<110> slip systems by planar dislocation glide; both $\frac{1}{2}\langle 110 \rangle$ type dislocations and stacking faults (SFs) have been observed suggesting the splitting of some perfect dislocations into Shockley partials ($\frac{1}{6}\langle 112 \rangle$ type)[11]. At cryogenic temperatures and at higher strains (~20%), nano-twinning becomes an additional deformation mechanism[1,11]. Twinning has also been observed after severe plastic deformation at room temperature by high-pressure torsion[18] as well as after rolling[12]. It has been argued that since the combination of dislocation slip and twinning provides a steady source of strain hardening to inhibit necking[1,10,11], ductility is enhanced together with strength to give excellent toughness (by intrinsic toughening)[1,5,19]. Similar arguments have been advanced to explain the extensive ductility of the so-called twinning-induced plasticity (TWIP) steels[20–22].

However, the detailed mechanisms and crack tip processes responsible for the high toughness of this high-entropy alloy have remained unclear. In this regard, understanding its defect behaviour, which is affected by such factors as the stacking-fault energy (SFE) and friction stress, is crucial. In this high-entropy alloy, the SFE is low and has been calculated to be ~20–25 mJ m$^{-2}$ (ref. 23). In addition, given the apparently random arrangement of its constituent atoms, which form a solid solution[24] down to the atomic scale[18,25], the alloy is expected to offer a relatively high resistance to dislocation motion (friction stress), especially at low homologous temperatures where the closely spaced barriers to dislocation motion (high concentrations of solute atoms) cannot be easily overcome by thermal activation[10,11,26].

In the present study, we attempt to identify the fundamental source of the excellent toughness of this high-entropy alloy at the nanoscale using in situ straining experiments in an aberration-corrected transmission electron microscope (TEM) to examine the microstructural evolution in the immediate vicinity of a crack. By imaging as close as within a few hundred nanometres of the crack tip, we identify multiple deformation mechanisms that are activated at different stages of deformation and act synergistically to contribute to the ultrahigh toughness. Deformability is initially afforded by the motion of the Shockley partial dislocations and the corresponding formation of SFs. However, as the applied stress increases, perfect dislocations start to move but their motion is observed to be extremely difficult. They move in localized bands containing arrays of many closely packed dislocations. These bands act as strong barriers for partial dislocation motion, which creates an outstanding strengthening effect. Strengthening is also caused by parallelepiped sessile volume defects formed by the interaction of partials slipping on different planes that block the motion of other dislocations. Finally, in the later stages of deformation, we see evidence of the creation of nanoscale bridges that span the crack-tip region and deform by nano-twinning, and as such are believed to be strong and ductile. As these nanobridges carry load that would be otherwise used to propagate the crack, they provide a potent means to inhibit crack advance (by extrinsic toughening[5]).

## Results

**Initial deformation modes.** Using in situ TEM straining experiments on the high-entropy CrMnFeCoNi alloy, performed at room temperature using a technique[27] of pulsed tensile loading of thin foil samples with electron-transparent regions containing cracks, we found that during the early stages of deformation, extensive Shockley partial ($\frac{1}{6}\langle 112 \rangle$ type) dislocation activity was evident near the crack tip. In particular, we observed slip of the leading partial dislocations, formation of corresponding SFs and glide of the trailing partial dislocations. Figure 1 shows images obtained during such in situ straining; the beam direction was [110] and the approximate location of the crack tip is marked by yellow lines (Fig. 1c,d). As deformation proceeds, numerous SFs on different {111} slip planes (indicated by the red arrows in Fig. 1a,b) were seen to continuously nucleate and annihilate in the vicinity of the crack tip indicating easy movement of the partial dislocations. The atomic-scale dynamic dislocation processes are shown in Fig. 1c,d. Fast Fourier transform patterns (inset in Fig. 1c) confirmed the zone axis of <110>, which allowed the direct imaging of the fcc stacking. After several displacement pulses, multiple partial dislocations nucleated and propagated on {111} slip planes, leaving behind SFs (Fig. 1e), as confirmed by the magnified fast Fourier transform image, which shows the shift from fcc to local hexagonal close-packed stacking sequence within the faults. A Burgers circuit was drawn within the fault and the closure failure was measured to be ~$\frac{1}{12}\langle 112 \rangle$. Subsequently, a trailing partial dislocation nucleated and propagated on the same slip plane as the foregoing partial, removing the stacking fault between them. (Estimates of the specific evolution of strain by these processes in the crack-tip region are given in Supplementary Note 1 and Supplementary Fig. 1). These dynamic dislocation sequences clearly demonstrate that because of the low SFE in this alloy[10,11,23], the slip of partial dislocations and the corresponding formation of SFs have a critical role in accommodating the strain at the early stages of plastic deformation. Real-time imaging of these sequences can be seen in Supplementary Movies 1 and 2.

The interactions of partial dislocations also generate other obstacles to dislocation motion that lead to further hardening. If slip on multiple {111} planes has been effectively activated, different partial dislocations may intersect each other to form 'stacking-fault parallelepipeds' through the interaction of three pairs of parallel {111} faults (Fig. 2a,b). Specifically, Fig. 2b shows

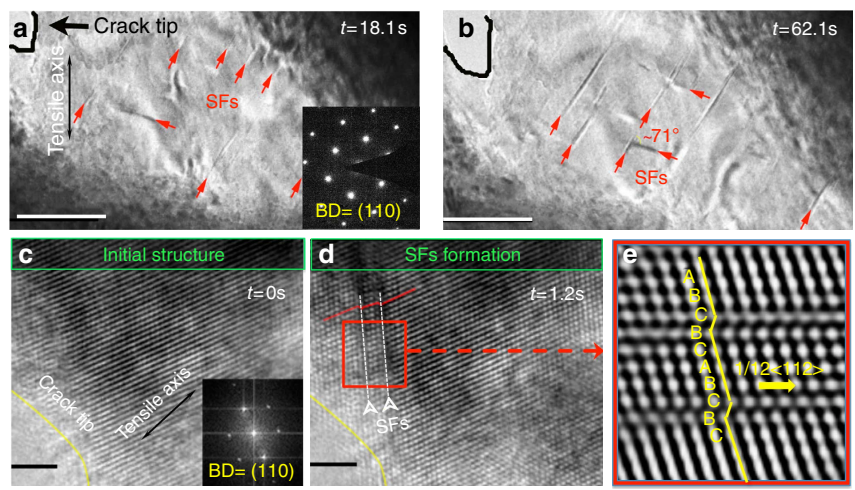

**Figure 1 | Partial dislocation activity and stacking-fault formation.** (**a**,**b**) Bright-field TEM images that show the formation of SFs (indicated by the red arrows) at the crack tip (top left-hand corner) under *in situ* loading of the CrMnFeCoNi high-entropy alloy (scale bar, 50 nm). Beam direction is [110]. (**c**,**d**) High-resolution TEM images captured from the *in situ* high-resolution TEM movie (scale bar, 2 nm). The formation of multiple SFs at the crack tip (bottom left-hand corner) was observed at the atomic scale. (**e**) is the magnified inverse fast Fourier transform image showing the atomic structure and stacking sequence (marked in yellow) of the SFs, surrounded by the red box in Fig. 1d. (The rapid motion of the Shockley partial dislocations and corresponding formation of SFs near the crack tip can be seen in real time in Supplementary Movies 1 and 2).

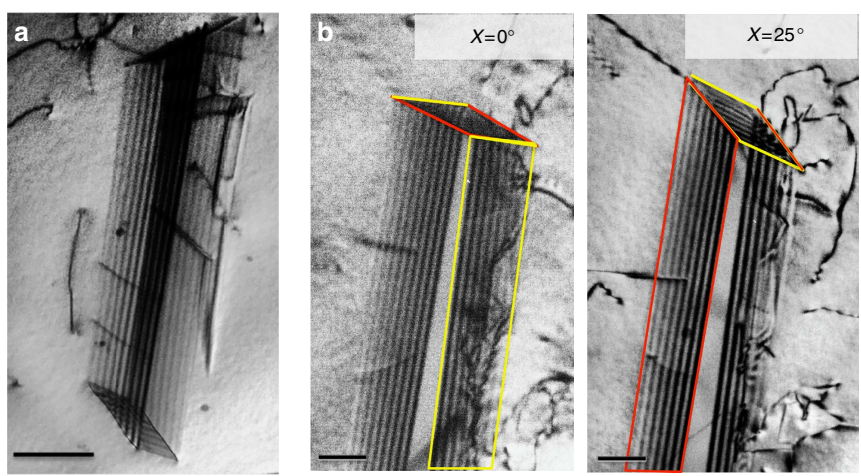

**Figure 2 | Stacking-fault parallelepipeds.** (**a**) TEM image of the stacking-fault parallelepiped structure (with faces lying on three sets of {111} planes; scale bar, 200 nm). (**b**) shows the same stacking-fault parallelepiped at different x-tilting angles (scale bar, 500 nm). When slip on multiple {111} planes has been effectively activated, different partial dislocations may intersect each other to form such stacking-fault parallelepipeds whose faces comprise three pairs of parallel {111} faults. These stable volume defects can lead to significant strain hardening; they present a formidable obstacle for other dislocations to overcome.

TEM images of the same stacking-fault parallelepiped viewed from different tilting angles. With this fault structure, two partial dislocation systems are activated, each of them on two parallel planes to form a dislocation rhomboid, which is a volume defect around which the tangling of dislocations occurs. Thus, although the movement of partial dislocations is relatively easy and provides initial deformability in this alloy, their interaction can lead to significant strain hardening through the formation of such three-dimensional stacking-fault defects that are both stable and a formidable obstacle for other dislocations to overcome.

**Subsequent deformation modes.** As the displacement applied by the straining stage increased, perfect dislocations started to move. In stark contrast to the relatively easy motion of the partials, the movement of perfect (undissociated, $\frac{1}{2}\langle 110 \rangle$ type) dislocations was observed to be quite difficult. Our *in situ* TEM observations demonstrate that the motion of these dislocations is not smooth; rather, they move in tiny segments with extremely low velocity. As a consequence, close-packed dislocation arrays form and the perfect dislocations move in localized bands leading to planar slip, as shown in real time in Supplementary Movie 3. Figure 3a shows a series of TEM images at different times of dislocations in a slip band in the crack-tip region captured from the *in situ* straining experiments showing that even when multiple displacement pulses were applied, the movement of the perfect dislocations was extremely slow. It is apparent that the motion of undissociated dislocations in this high-entropy alloy needs to overcome continuous activation barriers, which presumably accounts for the high friction stress. The localized bands of planar slip with slow-moving perfect dislocations act as barriers to, and significantly impede the motion of, the fast moving partial dislocations (as can be seen in real time in Supplementary Movie 4). The partial dislocations stopped in front of the bands of planar slip, resulting in a significant pile-up of partial dislocations. Figure 3b is a bright-field TEM image showing the interaction between partial dislocations and a localized band of planar slip, which blocks the partial dislocations and makes their subsequent motion more difficult. Additional force is needed for those partial dislocations to overcome the barrier. Extremely strong and complex dislocation interactions were observed. Such interactions are considered to be one of the important strengthening mechanisms during plastic deformation of this high-entropy alloy.

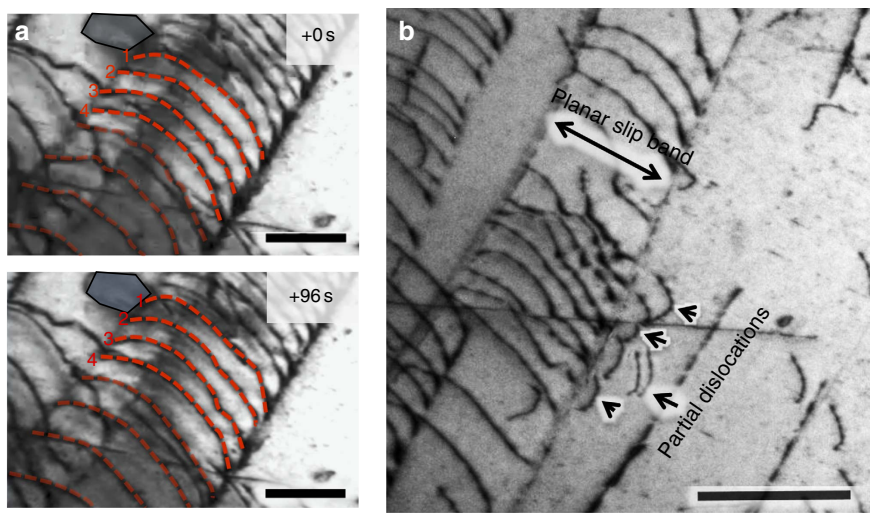

**Figure 3 | Slow planar slip of perfect dislocations.** (**a**) TEM images represent the dynamic process of the planar slip of undissociated $\frac{1}{2}\langle 110 \rangle$ type dislocations; their motion was observed to be slow and quite difficult (shown in real time in Supplementary Movie 3), in contrast to the easy motion of the $\frac{1}{6}\langle 112 \rangle$ type partials (shown in Fig. 1). Scale bar, 200 nm. (**b**) A bright-field TEM image showing the blocking of partial dislocations by the localized band of planar slip (scale bar, 500 nm). As shown in Supplementary Movie 4, partial dislocations move fast but can be abruptly arrested at the localized bands of planar slip containing arrays of many closely packed perfect dislocations. The strong interaction between them provides a significant hardening effect.

**Extrinsic toughening by crack-tip bridging.** In addition to identifying a sequence of deformation modes in this alloy that can explain its intrinsic toughness based on its inherent deformability and strain hardening, that is, its ductility and strength, our atomic-resolution imaging during *in situ* straining also revealed certain unusual mechanisms of extrinsic toughening for a metallic material in the immediate vicinity (within a few hundred nanometres) of the crack tip (Fig. 4). First, nanovoids were observed to form, in the absence of second-phase particles, in the 'semi-cohesive zone' in the vicinity of the crack tip, from the intersection of two {111} slip planes at the later stages of deformation. Interestingly, the formation of these nanoscale- to submicron-sized voids resulted in the creation of numerous nanoscale fiber-like regions that bridged the crack faces for about a few hundred nanometres behind the crack tip (Fig. 4a). Consequently, the opening of the crack was accompanied by many individual nanoscale deformation processes within these nanobridges. For example, a large proportion of these bridges displayed significant elongation when twinning shear was found to be the primary deformation mode that accommodated strain. Figure 4b shows the typical response to the tensile loading of the nanobridges spanning the crack; deformation twinning was observed during elongation of the bridge, as shown in the high-resolution TEM image of this region taken during the *in situ* test (Fig. 4c). The formation and twinning sequence in the nanobridges is also shown in Supplementary Fig. 2 and in real time in the Supplementary Movies 5 and 6.

As our *in situ* TEM testing was performed using a single-tilt holder, we conducted further evaluation of the atomic structure of the deformation twins using an aberration-corrected Titan TEM with a double-tilt holder. Figure 4d shows the representative post-mortem atomic structure of the nanoscaled deformation twins, marked by the red arrow in Fig. 4b and viewed along the [110] direction, revealing a typical {111} twin structure with a highly coherent twin boundary, characteristic of fcc metallic materials[28]. As analysed in the Supplementary Note 2, the energy dissipation achievable by twinning deformation can be much larger than by dislocation slip in the nanobridges. We believe that such deformation twinning-accommodated extension of the nanobridges spanning the crack provides a potent, yet unexpected, extrinsic toughening mechanism during the late stages of deformation, which impedes crack extension and further contributes to the excellent damage tolerance of this high-entropy alloy.

**Discussion**
Using *in situ* atomic/nanoscale experimental observations, we have found that the exceptional damage tolerance of the CrMnFeCoNi high-entropy alloy can be principally associated with a novel synergistic sequence of plastic deformation mechanisms at different stages of strain that are the source of its excellent combination of strength, ductility and toughness.

In general, toughness or fracture resistance can be considered to result from two classes of mechanisms: 'intrinsic toughening' mechanisms, which are primarily associated with plasticity (and the generation of ductility) and operate ahead of the crack tip to provide resistance to microstructural damage, and 'extrinsic toughening' mechanisms, which operate principally in the wake of the crack tip to inhibit fracture by 'shielding' the crack from the applied driving force[5]. Intrinsic toughening is most effective in ductile materials; it is inherent to the material and is effective in inhibiting both the initiation and growth of cracks. Extrinsic toughening, for example, crack bridging, conversely, is the primary source of toughening in brittle materials; it is dependent on the crack size and geometry, and are only effective in inhibiting crack growth[5]. One of the unique aspects of this high-entropy alloy is that it can generate fracture resistance from both these classes of mechanisms; the numerous plasticity and hardening mechanisms creating both strength and ductility by intrinsic means, and the near-tip crack bridging, which is extremely rare in a ductile metallic material, providing additional toughening extrinsically.

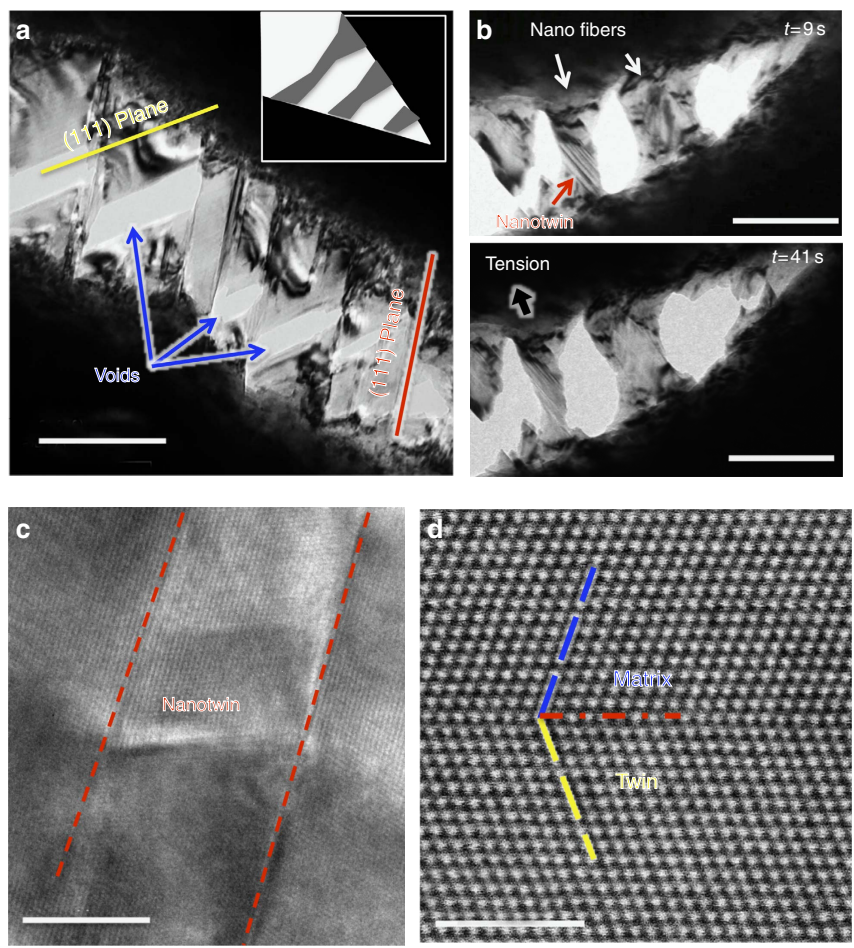

**Figure 4 | Crack bridging via near-tip twinned nanobridges.** (**a**) Bright-field TEM image of a growing crack during in situ straining of the CrMnFeCoNi high-entropy alloy, showing the formation of nano/submicron voids at the intersection of two slip systems, which then grow along two {111} slip bands (scale bar, 200 nm). The crack tip is located ~500 nm away from the right-lower corner of this image. The inset in **a** is a schematic of the structure at the crack tip. (**b**) two TEM images captured from the in situ TEM movie (Supplementary Movie 6) that show the nanoscale tensile loading of nano 'fibres' that bridge the crack in the near-tip region (scale bar, 200 nm). Nanotwins can be seen to form in some of the fibres, enhancing their ductility and resulting in their significant elongation. (**c**) High-resolutionTEM image of deformation twinning taken during in situ TEM tensile test (scale bar, 5 nm). (**d**) HAADF-STEM image of the atomic structure of the deformation nanotwins, showing characteristic {111} twins (scale bar, 2 nm).

Specifically, at the early stages of deformation, the easy motion of partial dislocations provides deformability and hence ductility to the material. At later stages, the mobility of partial dislocations is significantly affected since the slow motion of undissociated dislocations leads to the formation of localized bands of planar slip, which act as strong barriers for the motion of partial dislocations; moreover, dislocation barriers can result from the formation of stacking-fault parallelepipeds, that is, volume defects created by the interaction of the partials. Consequently, significant strain hardening is expected (that is significantly different from that in TWIP steels, where dislocation–dislocation and dislocation–twinning interactions are considered to be the major sources of hardening). Eventually, at the final stages of deformation, crack propagation appears to be impeded by the presence of twinned nanoscale fiber-like bridges across the crack, which act to shield the crack tip and provide a source of extrinsic toughening. We believe that the excellent strength, ductility and toughness of this high-entropy alloy results directly from this synergy of intrinsic (plasticity related) and extrinsic (shielding related) mechanisms[5], which (unlike traditional toughening mechanisms involving mainly precipitates and/or boundaries) primarily result from its special material characteristics of a low average SFE with possible variation of local SFE (that modulates the behaviour of full and partial dislocations locally)[10,11] and high lattice friction (that provides strengthening from the increased resistance to the planar slip of dislocations and its influence on the motion of partial dislocations), the low SFE further contributing to the remarkable formation of nanoscale crack bridges within a few hundred nanometres of the crack tip. (Further deliberations on the controlling factors for this synergy of deformation mechanisms are described in Supplementary Note 3).

In light of the unique nature of the origin of the tensile strength and ductility and the resulting excellent fracture toughness displayed by this high-entropy alloy, further optimization of these mechanisms may present new directions for the future design of advanced metallic materials with unprecedented damage tolerance.

## Methods

**Material processing and sample preparation.** The high-entropy CrMnFeCoNi alloy was processed, as described previously[1], by arc-melting high-purity elemental starting materials and drop casting into rectangular-cross-section copper moulds, followed by cold forging and cross rolling at room temperature into sheets roughly 10 mm thick. Recrystallization at 800 °C resulted in an equiaxed grain structure with an average grain size of ∼6 μm. Samples were prepared from the recrystallized sheet in the form of rectangular pieces (2.5 × 6.0 × 1.0 mm) cut out using electric discharge machining. They were then ground and mechanically polished with SiC papers down to a thickness of ∼80 μm. To produce electron-transparent regions for observation and analyses during the in situ TEM tensile tests, the mechanically polished specimens were further thinned using ion milling or jet polishing until a hole appeared in the middle of the foils. We used two different methods to rule out any influences from specific sample preparation methods. Ion milling was performed using a voltage of 5 kV. Twin jet polishing was conducted at ∼10 °C using a current/voltage of 13 mA/12 V.

The thinned high-entropy alloy foils were glued on stainless-steel substrates as shown schematically by the dashed horizontal rectangle in Supplementary Fig. 3. The substrate contained two circular holes for the loading pins of the straining stage and a narrow rectangular window in the center for transmission of the electron beam and to ensure that the high-entropy alloy film stretched (approximately) uniaxially across it with minimum rotation. This set-up is similar to that used in a previous study[29].

**Characterization methods.** The in situ TEM tensile tests were conducted at room temperature with a Gatan model 654 single-tilt straining holder, in a FEI Tecnai G2 F20 TEM operating at 200 kV. Only samples that were well attached to the substrate, about 12, free of contamination from the sample preparation procedure and did not rotate or bend during deformation were selected for detailed TEM investigations. The tensile loading was accomplished by applying intermittent displacement pulses manually through a trigger switch that activated a motor in the straining holder, resulting in an axial displacement rate of ∼1.0 μm s$^{-1}$ during each pulse. During the holding period between pulses, the specimen remained in a strained state at a fixed displacement. Time-resolved TEM and high-resolution TEM images of the regions of interest were recorded with a Gatan CCD camera at 10 frames per second. The ion milling and jet polishing processes produced many tiny cracks in the electron-transparent region around the hole, but our analyses focused mainly on the microstructure near those cracks that were roughly perpendicular to the loading direction. The regions in front of the crack tip were monitored during the straining process. The thickness of the TEM foils was not uniform; the regions closest to the hole were the thinnest. Since the thickness of the foil increases gradually, the cracks grew from the thin area (electron-transparent region) to the thick area (non-electron transparent). However, the plastic deformation of the crack tip area always made the tip region thinner than the area around it. Nanovoids and nanobridges then formed in the 'semi-cohesive' zone in the crack-tip region.

It should be noted here that such ultrahigh-resolution observations have to be made at the nanoscale to permit any characterization of mechanisms so close to the crack tip. Accordingly, thin foil samples must be used which, under load, cannot display the full constraint of a bulk sample.

### Acknowledgements
This work was supported in China by grants from the Chinese 1000-Youth-Talent Plan and the State Key Program for Basic Research of China (No. 2015CB659300) (for Z.J.Z., M.M.M., J.W., Z.Z., S.X.M. and Q.Y.) and in the US by US Department of Energy, Office of Science, Office of Basic Energy Sciences, Materials Sciences and Engineering Division, through the Materials Science and Technology Division at Oak Ridge National Laboratory (for E.P.G.) and the Mechanical Behaviour of Materials program (KC13) at the Lawrence Berkeley National Laboratory (for B.G. and R.O.R.).

### Author contributions
Q.Y. and R.O.R. conceived the study, Z.J.Z. and M.M.M. performed the experiments (directly supervised by J.W., Z.Z., S.X.M. and Q.Y.), B.G., E.P.G., R.O.R. and Q.Y. analysed the results and wrote the manuscript.

### Additional information
**Supplementary Information** accompanies this paper at http://www.nature.com/naturecommunications

**Competing financial interests:** The authors declare no competing financial interests.

**Reprints and permission** information is available online at http://npg.nature.com/reprintsandpermissions/

**How to cite this article:** Zhang, Z. et al. Nanoscale origins of the damage tolerance of the high-entropy alloy CrMnFeCoNi. Nat. Commun. 6:10143 doi: 10.1038/ncomms10143 (2015).